\documentclass[twocolumn]{article}
\advance\topmargin-1in\advance\oddsidemargin-1in
\evensidemargin\oddsidemargin

\makeatletter
\def\@normalsize{\@setsize\normalsize{12pt}\xpt\@xpt
\abovedisplayskip 10pt plus2pt minus5pt\belowdisplayskip \abovedisplayskip
\abovedisplayshortskip \z@ plus3pt\belowdisplayshortskip 6pt plus3pt
minus3pt\let\@listi\@listI}

\def\section{\@startsection {section}{1}{\z@}{20pt plus 2pt minus 2pt}
{8pt plus 2pt minus 2pt}{\centering\normalsize\sc
\edef\@svsec{\thesection.\ }}}
\def\thesection{\Roman{section}}

\def\subsection{\@startsection {subsection}{2}{\z@}{16pt plus 2pt minus 2pt}
{6pt plus 2pt minus 2pt}{\normalsize\sl
\edef\@svsec{\thesubsection.\ }}}
\def\thesubsection{\Alph{subsection}}

\long\def\@makecaption#1#2{
\vskip10pt\begin{center} #1 #2 \end{center}\par\vskip 1pt}
\def\fnum@figure{\raggedright{\footnotesize Fig. \thefigure }.%
\footnotesize}
\def\fnum@table{\footnotesize TABLE \thetable\\\footnotesize\sc}
\def\thetable{\Roman{table}}

\makeatother
\usepackage{booktabs}
\usepackage{algorithm}
\usepackage{algpseudocode}
\usepackage{amsmath}
\usepackage{amssymb}
\usepackage{array}
\usepackage{balance}
\usepackage[pangram]{blindtext}
\usepackage[noadjust]{cite}
\usepackage{caption}
\usepackage{color}
\usepackage{comment}
\usepackage{dblfloatfix}
\usepackage[inline]{enumitem}
\usepackage{epsfig}
\usepackage{etoolbox}
\usepackage{fancybox}
\usepackage{filecontents}
\usepackage{fixltx2e}
\usepackage{float}
\usepackage{graphicx}
\usepackage[bookmarks=false]{hyperref}
\usepackage{hyperref}
\usepackage{lastpage}
\usepackage{lipsum}
\usepackage{listings}
\usepackage{multirow}
\usepackage{multicol}
\usepackage{pifont}
\usepackage{placeins}
\usepackage{rotating}
\usepackage{setspace}
\usepackage{soul}
\usepackage[hang, tight]{subfigure}
\usepackage{threeparttable}
\usepackage{times}
\usepackage{url}
\usepackage{verbatim}
\usepackage{wrapfig}
\usepackage[usenames,dvipsnames]{xcolor}
\usepackage{color}

\newcolumntype{I}{!{\vrule width 3pt}}
\newlength\savedwidth

\newlength\savewidth

\algtext*{EndWhile}
\algtext*{EndIf}
\algtext*{EndFor}
\algtext*{EndProcedure}

\begin{document}
\date{}

\title{
\vspace{-0mm}
\Large\textbf{Efficient Neural Network Implementation with Quadratic Neuron}
\vspace{0mm}}	


\author{Zirui Xu$\dagger$, Jinjun Xiong$\ddagger$, Fuxun Yu$\dagger$, Xiang Chen$\dagger$\\
$\dagger$George Mason University, { $\{$zxu21, fyu2, xchen26$\}$@gmu.edu }\\
$\ddagger$IBM T. J. Watson Research Center, jinjun@us.ibm.com
\vspace{0mm}}



\maketitle
\thispagestyle{empty}


\newbool{inccomment}
\setbool{inccomment}{true}
\newcommand{\XX}[1]{\ifbool{inccomment}{{\color{magenta} #1}}{}}
\newcommand{\CT}[1]{\ifbool{inccomment}{{\color{magenta}CT\@: #1}}{}}
\newcommand{\NT}[1]{\ifbool{inccomment}{{\color{blue}NT\@: #1}}{}}
\newcommand{\TD}[1]{\ifbool{inccomment}{{\color{orange}#1}}{}}
\newcommand{\FN}[1]{\ifbool{inccomment}{{\color{OliveGreen}#1}}{}}
\newcommand{\GR}[1]{\ifbool{inccomment}{{\color{Tan}#1}}{}}
\newcommand{\LD}{\ifbool{inccomment}{{\color{magenta}\\============================================\\}}}
\newcommand{\RF}{\ifbool{inccomment}{{\color{green}~[R]}}}
\newcommand{\tabincell}[2]{\begin{tabular}{@{}#1@{}}#2\end{tabular}}
\newcommand{\roma}[1]{\uppercase\expandafter{\romannumeral #1\relax}}
\graphicspath{{}}

\vspace{-5mm}
\section{Introduction}
\label{sec:intr}
\vspace{-2mm}

In the past few decades, researchers have found that the neural networks work like a function approximation machine: By training in the machine learning applications, the neural networks learn to approximate the unknown underlying mapping function from inputs to outputs ~\cite{goodfellow2016deep}. Previous works proved that the combination of the linear neuron network with nonlinear activation functions (\textit{e.g.} ReLu) can achieve nonlinear function approximation ~\cite{fan2020universal}. By embedding massive linear neurons and activation functions, the nonlinearity of approximation can be enhanced theoretically. However, simply widening or deepening the network structure will introduce some training problems \cite{hochreiter1998vanishing,pascanu2013difficulty}.

Recently, some works investigated such nonlinear functional approximation from another angle: by directly modifying the first-order function in the original neuron to a second-order version, they upgraded the linear neuron to a quadratic one. 
Compared with the linear neuron, the quadratic neuron shows following two major advantages: 
1) Due to the second-order function inside, the quadratic neuron has a nonlinear functional approximation itself, which potentially eliminates the need for nonlinear activation functions.
Therefore, the structures of the second-order CNNs can be more unified and simplified, improving the training converge speed.
2) Higher approximation ability indicates that much more information is encoded into each quadratic neuron. Therefore, the second-order CNN shows higher generalization ability, directly benefiting the learning performance (\textit{e.g.} prediction accuracy).
However, most of the previous works still focused on the quadratic neuron design \cite{gupta2004static, fan2018new, jiang2019nonlinear}, 
ignoring the challenges and corresponding solutions during network practical deployment.

In this work, we are aiming to build a comprehensive second-order CNN implementation framework includes neuron/network design and system deployment optimization. 
Specifically, we first consider both computation-efficient and implementation-friendly requirements during the quadratic neuron design phase, decreasing the space and time complexity of the proposed quadratic neuron from $O(n^2)$ to $O(n)$.
Second, we identify the potential challenges during the second-order network deployment in two levels: compiler/library, and hardware resource. We propose the corresponding optimization solutions to enable the second-order CNN can achieve an optimal performance during deployment.
Finally, we conduct preliminary experiments to show the proposed second-order CNN design has better learning performance and converge speed. The results also provide the fundamental baselines for the further deployment optimization.
\vspace{-4mm}
\section{Implement-oriented Quadratic Neural Design}
\label{sec:algo}
\vspace{-1.5mm}





\begin{figure}[t]
	\centering
	\captionsetup{justification=centering}
	\vspace{-3mm}
	\includegraphics[width=3.8in]{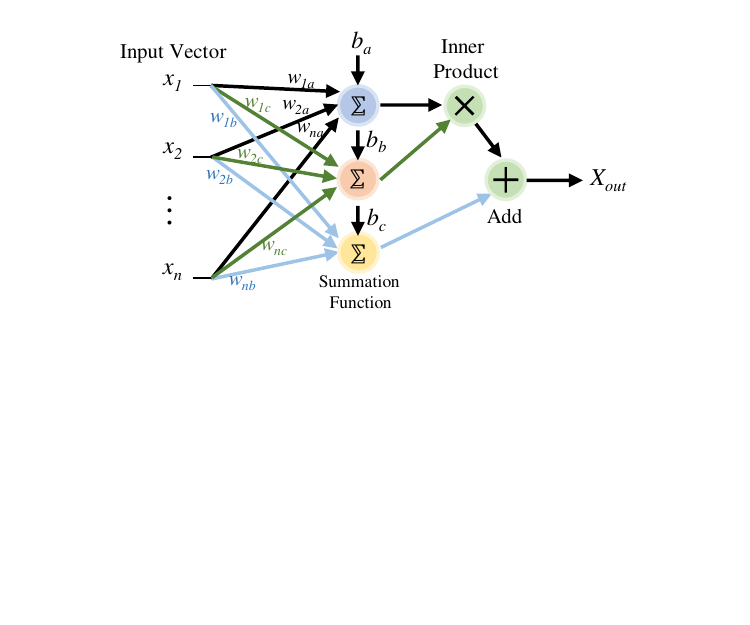}
	\vspace{-45mm}
	\caption{The Structure of the Proposed Quadratic Neuron}
	\vspace{-4mm}
	\label{quatratic_neuron}
\end{figure}





Recently, quadratic neuron drew more attentions from research community and several quadratic neuron formulations are proposed ~\cite{gupta2004static, fan2018new, jiang2019nonlinear}. 
The most popular one among them is formulated as:
\begin{equation}
\scriptsize
	X_{out} = \sum^{n}_{i=0}\sum^{n}_{j=0}w_{ij}x_i x_j = \mathbf{x}^T \mathbf{W_{\eta}} \mathbf{x}
	\label{eq:metric1}
	\vspace{-2.5mm}
\end{equation}
\normalsize
where $\mathbf{x}^T$ denotes the transposition of the input vector and $\mathbf{W_{\eta}}$ represents a $n \times n$ trainable weight matrix. 
	This quadratic neuron formulation has $O(n^2)$ time and space complexity, which is not practical for large neuron network with thousands even millions neurons. 
	Moreover, the implementation of such quadratic neuron design is difficult since the state-of-the-art deep learning frameworks can hardly support it. 

By considering the above two limitations, we propose a new quadratic neuron formulation, which is shown in Fig.~\ref{quatratic_neuron}. 
We decompose $\mathbf{W_{\eta}}$ into two weight vectors so the proposed quadratic neuron is composed of three ordinary linear neurons:
\begin{equation}
\scriptsize
	X_{out} = (\sum^{n}_{i=1}x_{ia} w_{ia} + b_a)(\sum^{n}_{i=1}x_{ib} w_{ib} + b_b)+(\sum^{n}_{i=1}x_{ic} w_{ic} + b_c), 
	\label{eq:metric1}
	\vspace{-2.5mm}
\end{equation}
\normalsize
where $w_{ia}$, $w_{ib}$, and $w_{ic}$ are trainable weights in three neurons. This quadratic neuron design addresses the above two problems, reducing both space and time complexity to $O(n)$ and can be easily implemented on the current deep learning frameworks such as \textit{Pytorch} and \textit{Tensorflow}. 
Moreover, compared with other recent quadratic neuron works \cite{fan2018new, jiang2019nonlinear}, our proposed neuron has better learning performance and can converge faster (will show in the experiment part). 


Since the proposed quadratic neuron demonstrates the strong nonlinearity of function approximation, we exclude the activation functions from the proposed second-order CNN structure.
The network without activation functions will has two characteristics: 1) any parameter in the network can directly derive its close-formed partial derivative, 
provides the insights for deployment optimization during the backward process.
2) it has computation workload reduction and faster converge speed (will show in the experiment part).

\vspace{-5mm}
\section{Second-Order CNN Deployment Optimization}
\label{sec:imag}
\vspace{-1mm}

Although we successfully upgraded the network structure to second-order, the current deep learning computing system is originally designed for the first-order CNNs. Therefore, several issues will be introduced in the following system levels due to this mismatch.

\textbf{Compiler/Library Level:} Currently, most of the network optimization functions provided by the deep learning libraries are first-order such as SGD ~\cite{bottou2010large}, which are specifically optimized for the first-order networks. On the contrary, our second-order CNN naturally achieves optimal converge speed under the second-order optimization methods. Therefore, we optimize the current libraries to enable them can support the second-order optimization methods. One specific technique we can use is adding Newton method (a representative second-order optimization method) and its corresponding Hessian matrix calculation function into optimization library (\textit{i.e.} the library that defines the optimizer).




\textbf{Hardware Resource Level:} During the network implementation on hardware, memory resource is always a main concern. Since a second-order CNN leverages more parameters in each neuron to achieve better nonlinear function approximation ability, directly applying current memory management scheme will bring higher memory occupancy rate. 
Therefore, we propose an optimized memory management scheme to reduce memory cost of the second-order CNN. By considering the fact that parameters have close-formed partial derivative expressions, the proposed scheme can only store part of essential parameters' gradients (\textit{e.g.} the output and input of each neuron) and calculate others based on the stored ones.

\vspace{-5mm}
\section{Preliminary Experiments}
\label{sec:Expe}
\vspace{-2mm}

In order to evaluate the second-order CNN's design superiority in terms of learning performance and converge speed, we conduct two preliminary experiments. The experiment results also provide a performance proof and a fundamental baseline for the deployment optimization as well.

	A multilayer perceptron and a ConveNet structure are evaluated on the MNIST \cite{lecun1998gradient} and CIFAR-10 \cite{krizhevsky2009learning} datasets, separately. The performance of the proposed second-order CNN is compared with a first-order CNN, and another second-order CNN proposed in \cite{fan2018new} (all of them have same neuron numbers). Table~\ref{tab:1} and Table~\ref{tab:2} show the comparison results. 
	From the tables we can find: 
1) First-order CNNs highly depend on activation functions to increase their nonlinearity of approximation abilities while second-order CNNs not. Actually, removing ``ReLu'' from our second-order CNN will make the network structure become more unified, improving converge speed at most 20\%.
2) Compared with the previous second-order CNN \cite{fan2018new}, our proposed CNN shows 5\% higher prediction accuracy with 2.5$\times$ faster converge speed. This is because our network is more robust during the training phase.
3) If optimization method is switched from SGD to L-BFGS (second order quasi-newton method), the training is speedup, showing the potential improvement introduced by library-level optimization. By applying other optimization strategies, both the epochs and time/epoch are expected to achieve further reduction.



\begin{table}[t]
	\vspace{0mm}
	\centering
	\caption{Evaluation of Multiple Perceptron on MNIST}
	\vspace{-2mm}
\footnotesize
\begin{tabular}{|c|c|c|c|c|}
\hline
\begin{tabular}[c]{@{}c@{}}Neuron Type\end{tabular} & ReLu & Accuracy & Epoch & Time/Epoch \\ \hline
First-order                                              & Yes  & 96.81\%  & 23    & 2.1s       \\ \hline
First-order                                              & No   & 92.07\%  & 9     & 2.1s       \\ \hline
\cite{fan2018new}                                               & Yes  & 97.37\%  & 9     & 2.3s       \\ \hline
\cite{fan2018new}                                               & No   & 97.31\%  & 7     & 2.3s       \\ \hline
Ours                                                  & Yes  & 97.25\%  & 9     & 2.2s       \\ \hline
Ours                                                  & No   & 97.47\%  & 8     & 2.2s       \\ \hline
Ours(L-BFGS)                                                  & No   & 97.66\%  & 4     & 3.9s       \\ \hline
\end{tabular}
\label{tab:1}
\end{table}
\normalsize
\vspace{-2mm}

\begin{table}[t]
	\vspace{-2mm}
	\centering
	\caption{Evaluation of ConvNet on CIFAR-10}
	\vspace{-2mm}
\footnotesize
\begin{tabular}{|c|c|c|c|c|}
\hline
Neuron Type & ReLu & Accuracy & Epoch & Time/Epoch \\ \hline
First-order                                              & Yes  & 76\%     & 135   & 2.7s       \\ \hline
First-order                                              & No   & 72\%     & 172   & 2.4s       \\ \hline
\cite{fan2018new}                                               & Yes  & 75\%     & 179   & 4.4s       \\ \hline
\cite{fan2018new}                                               & No   & 73\%     & 161   & 4.3s       \\ \hline
Ours                                                  & Yes  & 78\%     & 74    & 4.3s       \\ \hline
Ours                                                  & No   & 78\%     & 60    & 4.1s       \\ \hline
Ours(L-BFGS)                                                  & No   & 78\%  & 47     & 7.9s       \\ \hline
\end{tabular}
\label{tab:2}
\end{table}
\normalsize
\vspace{-0mm}

	\vspace{-2mm}
\section{Conclusion}
\label{sec:conc}
	\vspace{-1mm}

In this work, we proposed a comprehensive second-order CNN implementation framework which includes the new quadratic neuron design and system deployment optimization. 
The experiment results shows the superiority of the proposed second-order CNN than previous works \textit{w.r.t} learning performance and training converge speed.




\let\oldbibliography\thebibliography
\renewcommand{\thebibliography}[1]{%
  \oldbibliography{#1}%
  \setlength{\itemsep}{-5pt}%
}

\vspace{-3mm}
\scriptsize
\bibliographystyle{IEEEtran}
\bibliography{KDD}
\end{document}